\documentstyle{article}
\def\be{\begin{equation}}
\def\ee{\end{equation}}
\def\bea{\begin{eqnarray}}
\def\eea{\end{eqnarray}}

\def\Br{\Biggr}
\def\br{\biggr}
\def\Bl{\Biggl}
\def\bl{\biggl}
\def\l{\label}
\def\r{\ref}
\begin{document}

\hfill{}

\vskip 3
 \baselineskip
 \noindent
\begin{center}
 {\Large\bf Eigenvalue Spectrum of a  Dirac Particle
in Static and Spherical  Complex Potential}
\end{center}
\vskip \baselineskip

Khaled Saaidi{\footnote {{E-mail-1 :  KSaaidi@ipm.ir }
\\{E-mail-2: ksaaidi@uok.ac.ir }}}

\vskip\baselineskip

{\small Department of Science,  University of Kurdistan, Pasdaran
Ave., Sanandaj, Iran } \\

 {\small Institute for Studies in Theoretical Physics and Mathematics,
 P.O.Box, 19395-5531, Tehran, Iran}

\vskip 2
\baselineskip



\vskip 2\baselineskip

\begin{abstract}
 It has been observed that a quantum
 theory need not to be Hermitian to have a real
spectrum. We study the non-Hermitian  relativistic quantum
theories  for many complex potentials, and we obtain the real
relativistic energy eigenvalues  and corresponding eigenfunctions
of a Dirac charged particle in complex static and spherically
symmetric potentials. Complex Dirac-Eckart, complex
Dirac-Rosen-Morse II, complex Dirac-Scarf and complex
Dirac-Poschl-Teller potential  are investigated.
\end{abstract}

\newpage
\section{Introduction}
The first interest study in the non-Hermitian quantum theory date
back to an old paper by Caliceti et al \cite{cal}. The imaginary
cubic oscillator problem in the context of perturbation theory has
been studied in \cite{cal}. The energy spectrum of that model is
real and discrete. It shows that one may construct  many new
Hamiltonians that have real spectrum, although, their Hamiltonians
are not Hermitian. The key idea of the new formalism
(non-Hermitian quantum theory) lies in the empirical observation
that the existence of the real spectrum need not to necessarily be
attributed to the Hermiticity of the Hamiltonian.  Such
non-Hermitian formalism, for the context of Schr\"{o}dinger
Hamiltonian have been studied for several models in [1-25]. The
analysis of the related purely real spectra of energies has been
performed with different techniques. For example, resummations of
divergent perturbation series[1], delta expansions [2], WKB method
[6], functional analysis [8], pseudo-Hermitian methods [17, 21,
23], and complex Lee algebra [18-20]. Also, some explicit studies
of the Hermitian and non-Hermitian
 Hamiltonians have performed in the context of Dirac
Hamiltonian. For example, the solution of ordinary (Hermitian)
Dirac equation for Coulomb potential including its relativistic
bound state spectrum and wave function was investigated in
[22,23]. Also by adding off-diagonal real linear radial term to
the ordinary Dirac operator, the relativistic Dirac equation with
oscillator potential has been introduced [24, 25] then  the
energy spectrum of corresponding  eigenfunctions have been
obtained. The ordinary (Hermitian) Dirac equation for a charged
particle in static electromagnetic field, is studied for Morse
potential \cite{Alhaidari}. The non-Hermitian formalism, for the
context of Dirac Hamiltonian, have been studied for  three
dimensional complex Dirac-Morse and complex Dirac-Coulomb
potentials  in [30, 31].

In this paper, we consider the non-Hermitian Dirac Hamiltonian for
several complex potentials. We study  a charged particle  in
static and spherically symmetric four vector complex potentials.
By applying a unitary transformation to Dirac equation, we obtain
the second order Schr\"{o}dinger like equation, therefore
comparison with well-known non-relativistic problems is
transparent. Using correspondence between parameters of the two
problems (the Schr\"{o}dinger equation  and the Schr\"{o}dinger
like equation which is obtained after applying the unitary
transformation to Dirac equation for a  potential ) we obtain the
bound states spectrum and corresponding  eigenfunctions.

The scheme of this article is as following. In sec.2, we study the
non-Hermitian version of Dirac equation for a charged particle
with static and spherically symmetric potential, and by applying
a unitary transformation we obtain the proper gauge fixing
condition and Schr\"{o}dinger like   differential equation. In
sec.3, we discus the Dirac equation for complex Dirac-Eckart
potential and obtain the real energy spectrum and corresponding
eigenfunctions. In sec.4, we consider the  Dirac equation for a
complex Dirac-Rosen-Morse II potential then we  obtain the real
eigenvalue and its wave function. In sec. 5, and sec.6, we
studies the complex Dirac-Scarf and Dirac-Poschl-Teller
potentials respectively.  We obtain the relativistic energy
spectrum and correspondening wave function for  the upper
component of spinors.

\section{Preliminaries}
The Hamiltonian of a Dirac particle for a complex electromagnetic
field is ($c = \hbar = 1 $)
 \be\l{1} H = \hat{\alpha}. (\hat{p} -
e \hat{A}) + \hat{\beta}m + eV, \ee where the Dirac matrices
$\hat{\alpha} , \hat{\beta}$ have their usual meaning, and setting
$A_{0}$ equal to V. In (1), $\hat{A} $ and V are the vector and
scalar  complex field respectively, where, $\hat{A^*} \neq
\hat{A}$ and $V^* \neq V$. Then the Dirac Hamiltonian (1) is not
Hermitian. It is well known that the local gauge symmetry in
quantum electrodynamic implies an invariance under the
transformation as:
 \be\l{2}
 (V, \hat{A})\rightarrow (V' , \hat{A'}) = (V + \frac{\partial
\Lambda}{\partial t}, \hat{A}+ \overrightarrow{\nabla}\Lambda ).
\ee Here $\Lambda(t,\overrightarrow{r})$ is a complex scalar
field. Suppose that the charge distribution is static with
spherical symmetry,  so the gauge invariance implies that $V' = V$
and $\hat{A'} = \hat{r}A(r)$, where $\hat{r}$ is the radial unit
vector\cite{Alhaidari}. One can denoted the correspondence wave
function of (1)  as: \be\l{3} \Psi=\left (
\begin{array}{rr}
\Phi \\
\chi \\
\end{array}\right ).
\ee In this case one can obtain
 \bea\l{4}
  (m + eV - E_r )\Phi & =& i{\br [} \hat{\sigma} .
\overrightarrow{\nabla} - e (\hat{\sigma}.\hat{r})A(r){\br ]}\chi, \nonumber \\
 (eV - m- E_r)\chi & =& i {\bl [}\hat{\sigma}.\overrightarrow{\nabla} + e
(\hat{\sigma}.\hat{r}) A(r){\br ]}\Phi. \eea Here $\hat{\sigma}$'s
are the three Pauli spin matrices, $E_r$ is relativistic energy,
and we  replaced
$ie\hat{\sigma}.\hat{A}(-ie\hat{\sigma}.\hat{A})$ in
first(second) equation of (\r{4}) instead of
$e\hat{\sigma}.\hat{A}$, respectively.   Note that, because of the
spherical symmetry of the complex field, the angular-momentum
operator $\hat{J}$ and the parity operator, $\hat{P}$, commute
with the Hamiltonian and the two spinors $\Phi$ and $\chi$ have
opposite parity also. So the correspondence wave functions are
denoted by \bea\l{5}
\Phi &=&  ig(r) \Omega_{\kappa,\mu}(\vartheta , \varphi), \nonumber \\
\chi&=&f(r)\sigma_{r}\Omega_{-\kappa,\mu}(\vartheta , \varphi).
\eea It is seen that \bea\l{6}
 (\hat{\sigma}. \overrightarrow{\nabla} )
 ig(r) \Omega_{\kappa,\mu}(\vartheta , \varphi)&=&
 i\sigma_{r} \Omega_{\kappa,\mu}
 ( \partial_{r}+\frac{1}{r} + \frac{\kappa}{r})g (r),\\
 (\hat{\sigma}.\overrightarrow{\nabla})
 (f(r)\sigma_{r}\Omega_{-\kappa,\mu}(\vartheta , \varphi))&=&
 \sigma_{r}\Omega_{-\kappa,\mu}(\partial_{r} + \frac{1}{r}
 - \frac{\kappa}{r})f(r),
 \eea
where $\hat{\kappa}$ is the spin orbit coupling operator as:
\be\l{8}
  \hat{\kappa}=\hat{\sigma}.\hat{L}+\hbar I.
  \ee
 and we have used from
 \be\l{9}
 \hat{\kappa}\Omega _{\mp\kappa,\mu}(\vartheta , \varphi) =
  \pm \kappa\hbar \Omega_{\mp \kappa,\mu}(\vartheta , \varphi),
 \ee
 in which
\be\l{10} \kappa =\left \{ \begin{array}{rr}
\;\;\;\  {-(l+1)} = -(j + {1 \over 2}) \;\;\;\;\;\; {\rm for} \;\;\;\; j = l +{1 \over 2} \\
l =(j + {1 \over 2}) \;\;\;\;\;\; {\rm for} \;\;\;\; j = l -{1 \over 2} \\
\end{array}\right.
\ee Therefore by defining $u_{1}=g(r)/r$ , $u_{2}=f(r)/r$, we
obtain the following two component radial Dirac equation
\cite{Grainer} \bea\l{11}
 (m + eV - E_r) u_1(r)&=&(\partial_r-\frac{k}{r}-eA(r))u_{2}(r), \nonumber\\
 (eV - m - E_r)u_{2}(r)&=&-(\partial_{r}+\frac{k}{r}+ eA(r))u_{1}.
 \eea
Note that, $A(r)$ is a gauge field, which has a symmetry such as
(\r{2}), therefore, it  must be fixed. It is seen that fixing this
gauge degree of freedom by  $\overrightarrow{\nabla}.
\overrightarrow{A}\equiv \frac{\partial A}{\partial r}= 0$ is not
a suitable choice. Remark that in this paper, instead of solving
Dirac equation we want to solve the $2^{th}$ order differential
equation, which is obtained by eliminating one component of
equation (\r{11}). However, for the cases which $eV \neq 0$, the
second order  differential equation is not Schr\"{o}dinger like
and, therefore,  one can obtain the proper gauge fixing by
applying the global unitary transformation on two components
$u_{1}$ and $u_{2}$ such as:
 \be\l{12} U=\left ( \begin{array}{rr}
a & ib \\
ib & a \\
\end{array}\right ),
\ee where $ a , b \in \Re $, and $a^2 + b^2 = 1 $. By applying
(\r{12} ) to the upper component, $ \phi^{u} $,  and lower
component, $ \phi^{l} $ of spinor  and institute them in (\r{11}),
we have
 \bea\l{13}
 (m -E_rC)\phi^{u} + {\bl [}\frac{i( S^{2}
- C^{2})}{S} eV - iSE_r - \partial_{r}{\br ]}\phi^l& =&
0, \nonumber \\
{\bl [}\frac{i(S^{2}- C^{2})}{S} eV - iSE_r + \partial_{r}{\br ]}
\phi^{u} - (m + E_r C) \phi^{l} &=& 0, \eea
 where, S= 2ab, $C= a^2 -b^2$ and  we have  used from  a gauge fixing
condition as: \be\l{14} eV = \frac{iS}{C}(eA + \frac{k}{r}). \ee

However, we eliminate the $\phi^{l}$ component in (\r{13}), and
obtain the radial differential equation for $\phi^{u}$ as:
\be\l{15}
 -\frac{d^{2}\phi^{u}}{dr^{2}} + V_{eff} \phi^{u}
 + (m^{2} - E_r^{2}) \phi^{u} = 0, \ee
where
 \be\l{16}
 V_{eff} = -\frac{(S^2-C^2)^2}{S^2}(eV)^{2} + 2E_r(S^2-C^2)(e V) - i\frac{(S^2-C^2)}{S}\frac{d(eV)}{dr}.\ee

Furthermore, from (\r{11}),  it is easily seen that for the
cases  which $eV = 0$ the unitary transformation is not
necessary. So, we can rewrite $u_1$ and $u_2$ as:
 \bea\l{121}
u_1&=&\phi^u, \nonumber \\
u_2 &=& \phi^l, \nonumber \eea  then, by eliminating $\phi^{l}$,
one can obtain the Schr\"{o}dinger like differential equation for
radial upper component, $\phi^{u}$, as:
 \be\l{17}
 -\frac{d^{2}\phi^{u}}{dr^{2}} + V_{eff} \phi^{u}
 + (m^{2} - E_r^{2}) \phi^{u} = 0, \ee
where
 \be\l{18}
 V_{eff} = (eA(r) + \frac{\kappa}{r})^{2} - \frac{d}{dr}(eA(r) + \frac{\kappa}{r}).\ee

\section{The complex Dirac-Eckart potential }

The complex Eckart potential which holds discrete energy spectrum
is \cite{znojil2} \be\l{19}
 V^{CE}(x)= \frac{A(A-1)}{\sinh(x)} -2iB\coth(x),
 \ee
 and the corresponding Schr\"{o}dinger equation is:
 \be\l{20}
{\bl (} -\frac{d^{2}}{dx^{2}} +\frac{A(A-1)}{\sinh(x)}
-2iB\coth(x) - E{\br )}\psi(x) =0. \ee Here $A, B \in\Re$. It is
obviously  seen that, the Eckart potential is singular in origin,
but, one can simply avoid their singularities by a local
deformation of the integration path. In \cite{znojil2}   by
solving the Schr\"{o}dinger equation (\r{20} ), the real energy
spectrum and corresponding wave functions were found as: \be\l{21}
  E_{n} = \frac{B^2}{(A-n)^2} - (A-n)^2 ,\hspace{1cm} n =
  1,2,...,n_{max}< A,
  \ee
and
 \be\l{22}
   \psi_n(x)= {\cal N}_n(\coth(x) -1 )^{\mu}(\coth(x) +1
   )^{\nu}P_n^{(2\mu, 2\nu)}(\coth(x)),
\ee where ${\cal N}_n$ is a normalization constant, $P_n^{(2\mu,
2\nu)}$ is Jacobi polynomial \footnote{ $P_n^{(\mu, \nu)}(x)$ is
Jacobi polynomial in all of this text  }and \bea\l{23} 2\mu
&=& (A-n), \nonumber \\
2\nu&=& -\frac{iB}{A-n}. \eea

Now by defining the  complex Dirac-Eckart four vector  potential
as: \be\l{24}
 (eV(r), eA(r)\hat{r}) = {\bl (}i\zeta\coth(r),
 (\frac{C\zeta}{S}\coth(r) - \frac{\kappa}{r})\hat{r}{\br )},
 \ee
  and by using (\r{16}), we can obtain,
   \be\l{25}
    V_{eff} = \frac{\eta(\eta -1)}{\coth^2(r)}- i\gamma\coth(r)-
    \eta^2,
    \ee
    where $\zeta \in \Re , \eta = \frac{\zeta}{S}(S^2-C^2)$ and $\gamma = 2E_r(C^2
    -S^2)\zeta$. So that the $2^{th}$ order differential equation
    for  radial upper component is:
    \be\l{26}
{\bl (}  -\frac{d^2}{dr^2 } + \frac{\eta(\eta -1)}{\coth^2(r)}-
i\gamma\coth(r)- (E_r^2 + \eta^2 -m^2){\br )}\phi^u_n(r) = 0. \ee

 Comparing (\r{26}) with (\r{20}) and then  using  (\r{21}) and
 (\r{22}), one can arrive at the relativistic real energy
 eigenvalue as:
 \be\l{27}
 E_{rn} = {\bl [} \frac{[m^2 - \eta^2 -(\eta-n)^2](\eta-n)^2}{[(1-2S)\eta -n][(1+2S)\eta
 -n]}{\br ]}^{1 \over 2},
 \ee
 in which
 \be\l{28}
 |n_{max} - \eta|< \sqrt{\eta^2 - m^2},
 \ee
 and the correspondence  wave function for $\phi^u_n(r)$ is the same
 eigenfunction in (\r{22}) with a new set of parameters  as:
 \bea\l{29}
 2\mu &=& (\eta-n), \nonumber \\
2\nu &=& - \frac{i\gamma}{\eta -n } = - \frac{i\gamma}{2\mu } \eea

\section{The complex Dirac- Rosen-Morse II potential}

The Schr\"{o}dinger equation for complex Rosen-Morse II potential
is:
 \be\l{30}
  -\frac{d^2\psi(x)}{dx^2} + V^{CRM}(x) \psi(x) -
E\psi(x) =0,
 \ee
where \be\l{31} V^{CRM}(x) = [(b_R + ib_I)^2 +q^2
-1/4]\textrm{cosch}^2(x)
 -2q(b_r +ib_I)\textrm{cosch}(x)\coth(x).
 \ee
 One can rewrite (\r{30}) as:
\be\l{32}
 -\frac{d^2\psi_n(x)}{dx^2} + V^{CRM}(x) \psi_n(x) =-(q-n-1/2)^2\psi_n(x) =0,
 \ee
which  shows that
\be\l{33}
E_n = -(q-n-1/2)^2,
\ee
where
\be\l{34}
n = 0, 1, 2,... n_{max} < q-1/2,
\ee
and also $q> 1/2$.
Hence, we define the Dirac-Rosen-Morse II four vector potential as:
\be\l{35}
 (eV(r), eA(r)\hat{r}) = {\bl (}0,
 [\zeta\coth(r) - (\eta_R+i\eta_I)\textrm{cosch}(r)
  - \frac{\kappa}{r}]\hat{r}{\br )}.
 \ee
Here, $\zeta, \eta_R, \eta_I \in \Re$. By making use of (\r{18}),
we find the effective complex Dirac-Rosen-Morse II  potential,
$V_{eff}^{CDRM}(r)$, as: \be\l{36} V_{eff}^{CDRM}(r) = {\bl
[}(\eta_R + i\eta_I)^2 + (\zeta +1/2)^2 -1/4){\br ]}
\textrm{cosch}^2(r) -2 (\zeta +1/2) (\eta_R + i\eta_I)
\textrm{cosch}(r)\coth(r) + \zeta^2. \ee Using (\r{36} ), we
obtain the following second order
 differential equation for the upper component, $\phi^u(r)$,
 \bea\l{37}
 {\bl [}-\frac{d^2}{dr^2}&+&[(\eta_R + i\eta_I)^2 + (\zeta +1/2)^2 -1/4)
]\textrm{cosch}^2(r)\nonumber \\
 & -&2 (\zeta +1/2) (\eta_R +
i\eta_I)\textrm{cosch}(r)\coth(r) {\br ]}\phi^u_n(r) =
   (E_{rn}^2 -m^2-\zeta^2)\phi^u_n(r). \eea
    Comparing (\r{37}) with
Schr\"{o}dinger equation for complex Rosen-Morse  II potential,
(\r{31}), gives the following real relativistic energy spectrum
for complex Dirac-Rosen-Morse II potential as: \be\l{38} E_{rn} =
\sqrt{m^2 + \zeta^2 - (\zeta-n)^2)}, \ee where $n = 0,1,2,...,
n_{max}.$ Here, $\zeta>0$, and reality of energy spectrum emphasis
that $n_{max}$ satisfy \be\l{39} |n_{max} - \zeta| < \sqrt{m^2
+\zeta^2}. \ee
\section{The complex Dirac-Scarf potential}
In \cite{znojil2}, the complex Scarf potential  and the
corresponding Schr\"{o}dinger equation is given in the form
\be\l{40} V^{CS}(x) = [(b_R + ib_I)^2 -q^2 +1/4]
\textrm{sech}^2(x)
 -2q(b_r +ib_I)\textrm{sech}(x)\tanh(x),
 \ee

 \be\l{41}
 {\Bl [} -\frac{d^2}{dx^2} + [(b_R + ib_I)^2 -q^2 +1/4]\textrm{sech}^2(x)
 -2q(b_r +ib_I)\textrm{sech}(x)\tanh(x) - E {\Br ]}\psi(x) =0.
 \ee
It is well known that the Schr\"{o}dinger equation of this potential
is exactly solvable.
 It is easily seen that this complex potential (\r{40}) is not
 invariant under PT-symmetry, but, for the cases which $b_R = 0$,
 is PT-symmetry invariant (where  P denotes the parity operator;
({P}$\psi)(x) = \psi(-x)$  and T the time reversal operator;
({T}$\psi)(x) = \psi^*(x) $  ).  For $q>1/2$, the associated
 eigenvalues and eigenfunctions are
\be\l{42} E_n = -(q-n-1/2)^2,\hspace{2cm} n = 0, 1, 2,... n_{max}
< q-1/2, \ee \be\l{43} \psi_n(x) = {\cal N}_n
\textrm{sech}^{q-1/2}(x)e^{ib_I\arctan(\sinh(x))} P_n^{(-b_I-q,
b_I-q)}(i\sinh(x)). \ee Therefore,  we define the complex
Dirac-Scarf four vector potential as: \be\l{44}
 (eV(r), eA(r)\hat{r}) = {\bl (}0,
 (\zeta\tanh(r) - (\eta_R+i\eta_I)\textrm{sech}(r)
  - \frac{\kappa}{r})\hat{r}{\br )}.
 \ee
Here, $\zeta, \eta_R, \eta_I \in \Re$. By making use of (\r{18}),
we find the effective complex Dirac-Scarf potential,
$V_{eff}^{CDS}(r)$, as: \be\l{45} V_{eff}^{CDS}(r) = {\bl
[}(\eta_R + i\eta_I)^2 - (\zeta +1/2)^2 +1/4){\br ]}
\textrm{sech}^2(r) -2 (\zeta +1/2) (\eta_R + i\eta_I)
\textrm{sech}(r)\tanh(r) + \zeta^2. \ee By substituting (\r{45} )
in (\r{18}), we have:
 \bea\l{46}
 {\bl [}-\frac{d^2}{dr^2}&+&
[(\eta_R + i\eta_I)^2 - (\zeta +1/2)^2 +1/4)
]\textrm{sech}^2(r)\nonumber\\
 &-&2 (\zeta +1/2) (\eta_R +
i\eta_I)\textrm{sech}(r)\tanh(r) {\br ]}\phi^u_n(r)
 = (E_{rn}^2 -m^2-\zeta^2)\phi^u_n(r). \eea Comparing (\r{46})
with Schr\"{o}dinger equation for complex Scarf potential,
(\r{40}),  gives the following real relativistic energy spectrum
for complex Dirac-Scarf potential as: \be\l{47} E_{rn}^2
-m^2-\zeta^2 = - (\zeta - n)^2, \ee and the corresponding
eigenfunction as: \be\l{48} \psi_n(r) = {\cal N}_n
\textrm{sech}^{\zeta}(r)e^{i\eta_I \arctan(\sinh(r))}
P_n^{(-\eta_I-q, \eta_I-q)}(i\sinh(r)). \ee Therefore, for $\zeta
>0$ the real relativistic bound state energy eigenvalue is:
\be\l{49} E_{rn} = \sqrt{m^2 + \zeta^2 - (\zeta-n)^2)},
\hspace{2cm} n = 0,1,2,..., n_{max}, \ee where, $n_{max}$ is the
largest positive integer which satisfy \be\l{50} |n_{max} -
\zeta| < \sqrt{m^2 +\zeta^2}. \ee
\section{The complex Dirac-Poschl-Teller Potential}
The complex Poschl-Teller Potential is \cite{znojil2} \be\l{51}
V^{CPT}(x) = \frac{M(M-1)}{\sinh^2(t)} -
\frac{N(N+1)}{\cosh^2(t)}, \ee where $t = x-i\epsilon$, $x \in
(-\infty , +\infty)$,
 $\epsilon \in(0, \pi/2)$ and $M, N \in \Re$. In this case, $V^{CPT}(x)$  is not singular at
 $x =0$. In \cite{znojil2}, the real eigenvalues and corresponding
  eigenfunctions of the  potential (\r{51}) were found by solving
  the Schr\"{o}dinger equation. It is found
  \be\l{52}
  E_n =-(2n +\sigma N + \tau M +(\tau - \sigma)/2)^2, \hspace{2cm}
  n=1, 2, ..., n_{max}<-\frac{1}{2}(\sigma N + \tau M +(\tau - \sigma)/2),
  \ee
  where $\tau = \sigma= \pm 1$ and
   \be\l{53}
   \psi_n(x) = {\cal N}_n\sinh^{\tau M}(t)\cosh^{\sigma N +1}(t)
   P_n^{(\tau M-1/2,\sigma N+1/2)}[\cosh(2t)].
   \ee
So, by defining the complex four vector as: \be\l{54}
 (eV(r), eA(r)\hat{r}) = {\bl (}0,
 (\zeta\tanh(t) - \eta\coth(t) - \frac{\kappa}{t})\hat{r}{\br )},
 \ee
where $t = r- i\epsilon , r\in [0, \infty), \epsilon \in(0,
\pi/2)$ and $\zeta$ and $\eta $ are real. We can obtain
$V_{eff}^{CDPT}$ as: \be\l{55} V_{eff}^{CDPT}(r) =
-\zeta(\zeta+1) \textrm{sech}^2(t) +
\eta(\eta+1)\textrm{cosch}^2(t) + (\zeta -\eta)^2.
 \ee
Therefore, the second order differential equation for the upper
spinor component of Dirac equation for Complex Dirac-Poschl-Teller is:
\be\l{56}
 {\bl [}-\frac{d^2}{dr^2} +
\frac{\eta(\eta+1)}{\sinh^2(t)} -\frac{\zeta(\zeta +1)}{\cosh^2(t)}-
(E_{rn}^2 -m^2-(\zeta- \eta)^2{\br ]}\phi^u_n(t)=0.
\ee
By comparing(\r{56}) with  the associated Schr\"{o}dinger equation of
the complex Poschl-Teller potential, (\r{51}), we can obtain the
relativistic real energy spectrum as :
\be\l{57}
  E_{rn} = \sqrt{m^2-(\zeta- \eta)^2 - (2n +\sigma \zeta + \tau \eta +(\tau - \sigma)/2)^2},
\ee where $  n=1, 2, ..., n_{max}$, which, $n_{max}$ satisfy
 \be\l{58}
2n_{max}<\sqrt{m^2 -(\zeta-\eta)^2}-(\sigma\zeta + \tau\eta
 +(\tau - \sigma)/2),
  \ee
and also the upper spinor component, $\phi^u_n(r)$ , is:
   \be\l{59}
   \phi^u_n(r) = {\cal N}_n\sinh^{\tau\eta }(t)\cosh^{\sigma\zeta +1}(t)
   P_n^{(\tau\eta-1/2,\sigma\zeta+1/2)}[\cosh(2t)],
   \ee
where ${\cal N}_n$ is normalization constant and $P^{(\mu, \nu)}_n$ is Jacobi
polynomial.

\section{Conclusion}
The non-Hermitian quantum theories have been studied for many
complex potentials. It is observed that a relativistic quantum
theory need not to be Hermitian to have a real spectrum. In this
paper we obtain the real relativistic energy eigenvalues of a
Dirac charged particle in  complex static and spherically
symmetric potentials.  We show that these complex Dirac potentials
have exact solution for all value of $\kappa$($\kappa$ is angular
momentum quantum number).

\end{document}